\documentclass[aps,pra,twocolumn,groupedaddress,showpacs]{revtex4-1}

\usepackage{graphicx}

\begin{document}

\title{Origin of the Norton-type wave scattered by a sub-wavelength metallic slit}

\author{J\'er\^ome Le Perchec}
\email[]{jerome.le-perchec@cea.fr}

\affiliation{C.E.A., LETI, Optronics Department, 17, avenue des Martyrs, 38054 Grenoble, France}


\begin{abstract}
We clarify analytically and numerically the physical origin and the behavior of the Norton field scattered
by a narrow slit, at optical frequencies. This apparent surface field, which comes in addition to the surface
plasmon-polariton and classic cylindrical light waves, features its own radiation lobe associated with oscillating
induced currents that spread over both horizontal metallic parts forming the slit. Theory is given taking into
account the finite size of the aperture and is illustrated with materials such as gold and amorphous silicon in
different spectral regions.
\end{abstract}


\maketitle

\section{Introduction}

A famous experiment \cite{ebbesen1998} reporting an abnormal light transmission (a few percent) through subwavelength hole arrays, sparked off a great surge of interest in the optical properties of nano-structured metallic surfaces. The authors initially advocated the unique role of surface plasmons. It is true that such modes are significantly present in the visible range, on noble metals. Nevertheless, subwavelength apertures, intrinsically, behave also as dipole antennas that are supposed to scatter a continuum of evanescent waves and quasi-omnidirectional space ones. In particular, these light waves may be preponderant at the close vicinity of the apertures, and can participate quantitatively to extraordinary transmission \cite{lalanne2008}. In other cases, taking narrow slits, dipolar interactions are responsible for some local enhancements and far-field modulations \cite{leperchec2006,leperchec2011}. Following many debates \cite{visser2006,weiner2009} about the actual mediation of extraordinary transmission at \textit{optical} frequencies (plasmon, surface lightwaves, or a mixture of them, depending on spectral window and geometry), efforts were done to finely describe analytically the electromagnetic field scattered by a slit, often reduced as a punctual source \cite{lalanne2009,nikitin2009,dai2009,gravel2012}. It is difficult to find a closed-form expression when the (complex) metal permittivity is finite, as known in the antennas context \cite{collin2004}. An intriguing result is the change of spatial damping of the non-plasmonic contribution far from the source, at the metal level \cite{lalanne2009}, which is reminiscent of a Norton-type wave \cite{nikitin2009} i.e. a ground radio wave \cite{wait1998}. One could believe that a kind of surface field with the wavevector of \textit{light} is also launched along the metal, together with the polariton, as a companion-wave that exchanges energy, but this terminology is not really appropriated, as we will see.

In most papers interested in this question, a purely mathematical approach of the scattering integral, strictly at the surface level, and with a punctual scatterer, is only partly satisfactory as we miss essential features to have a complete picture of the electromagnetic entity we actually consider. An analysis discriminating each contribution, in the broad space, and giving an explicit physical interpretation, is still lacking. Thus, this paper aims to refresh our vision in a practical way, and shows that the lateral Norton wave generated by a slit constitutes a \textit{third} electromagnetic contribution taking birth in the whole horizontal conducting surface. As this radiating subfield combines with the conventional dipole field in a special manner, it results in a weak apparent surface wave, however fundamentally different from a surface polariton. We will exemplify theoretical results essentially with gold in the infrared, but for sake of generality, the case of amorphous silicon (aSi) in the ultraviolet region will be sometimes illustrated since this material may also exhibit a metal-like behaviour (with strong absorbing properties).

\begin{figure}[h]
\includegraphics[angle=0,scale=0.38]{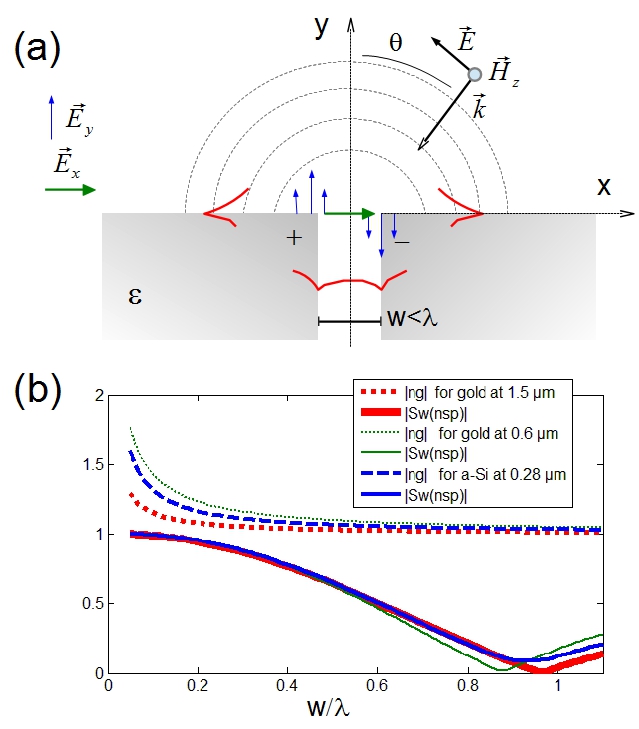}
\caption{(a) Sketch of the 1D sub-wavelength slit with its waveguide mode. When excited by a TM-plane wave, it scatters surface plasmons (in red) and space lightwaves (dotted lines). Arrows indicates the typical orientation of the electric field components just above the aperture. (b) Modulus of the index $n_g$ of the guided mode (\ref{ng}) and that of the aperture function $S_w(u=n_{sp})$ (\ref{Su}) for the polariton wave (\ref{Hsp}), depending on the slit width, and for three wavelengths ($\lambda=1.5, 0.6, 0.28 \mu$m), taking either the permittivity of gold, either that of amorphous silicon \cite{palik}. Let us recall that $\varepsilon_{aSi}=-3.7+13.8i$ at $\lambda=282$nm.}\label{figure1}
\end{figure} 

\section{Theoretical framework: from the perfect metal case to the real one}

As a tutorial case, we revisit the problem in the framework of the simplified, highly instructive modal method, by taking into account the fundamental waveguide mode inside the sub-$\lambda$ one-dimensional (1D) slit, in TM polarization (see Fig.\ref{figure1}(a)). In this paper, we \textit{do not} consider the indentation as a punctual scatterer. 

Take a material with a permittivity $\varepsilon$ strongly negative. The surface impedance boundary conditions are applicable while tangential wave vectors of the scattered waves are much smaller than $k|\varepsilon|$, i.e.,

\begin{equation} k_{\perp}= [ k^2\varepsilon-k_{//}^2 ]^{1/2}\approx k\sqrt{\varepsilon}, \label{kperp}  \end{equation}
 in the metal, where $k=2\pi/\lambda$. This is widely the case for noble metals in the infrared region and for reasonably sub-wavelength geometries. We will note $Z=\varepsilon^{-1/2}$ the surface impedance (small and essentially imaginary). After lengthy algebra \cite{leperchec2006}, and omitting the time dependence $e^{-i\omega t}$, we get the following sequel (for $y>0$):

\begin{eqnarray}
H_z(x,y)=[e^{-iky\cos\theta }+\frac{\cos\theta-Z}{\cos\theta+Z}e^{iky\cos\theta}]e^{ik\sin\theta x}  \nonumber \\
+\ \alpha(k,Z) \int_{-\infty}^{\infty}\frac{S_{w,Z}(u)}{v+Z}\ e^{ik(ux+vy)}du \label{Htot}\end{eqnarray}


where the vector $v=\sqrt{1-u^2}$ ($\arg(v)\in [0;\pi]$), and
\begin{equation}\label{Su} S_{w,Z}(u)=\frac{1}{2}[\sec((\nu+u)\frac{kw}{2})
 +\sec((\nu-u)\frac{kw}{2})] \end{equation}

with $\sec(x)=\sin(x)/x$ and $\nu=\sqrt{1-n_g^2}$. The effective index of a fundamental mode which is vertically guided along the slit is:

\begin{equation}\label{ng}
n_g=(1+\frac{2iZ}{kw})^{1/2}.
\end{equation}
This mode is built by the antisymmetric coupling of wall plasmons. $S_w(u)$ is the Fourier transform of its eigenfunction, and $\alpha$ its excitation coefficient \cite{leperchec2006} ($\alpha\propto E^{slit}_x(y=0)$ but it has no importance in the forthcoming discussion). The scattering integral (\ref{Htot}), that is to say the \textit{field structure}, is what interests us in this paper, and is \textit{independent} of the slit reaction. It is worth recalling that
$E_x(x,y=0) \propto Z\cdot H_z(x,y=0)$ at air/metal interfaces. 

First, let us briefly comment the perfect metal case. When $Z=0$ inside the slit ($\varepsilon=-\infty$, $n_g=1$), we immediately get $S_w(u)=\sec(kwu/2)$. And if the metal is perfect everywhere, the scattered field may be exactly\cite{wirgin1973} turned into an integral of some zero order Hankel function of the first kind $H_0^{(1)}$ over the slit width (see Appendix A for more details). Consequently, the field scattered in any direction is:
\begin{equation}\label{Hperf} 
H_z(kr>1) \approx \alpha (k) \sqrt{\frac{2\pi}{kr}}\ e^{i(kr-\pi /4)} 
\end{equation}
with $r=\sqrt{x^2+y^2}$ and $kw<1$. This solution verifies the Sommerfeld radiation condition. It is a \textit{cylindrical}, dipole-type field. Besides, it is possible to show that $E_y$ has an almost identical expression, so that the power flux $\mathbf{E_y}\times \mathbf{H_z^*}$ propagating along the perfect metal surface has a \textit{$1/x$ spatial damping}, for $kx>1$.

Let us come back to the real metal case and put aside the specular term in (\ref{Htot}). We know the scattered field (\ref{Htot}) is the sum of two main contributions: a surface polariton (SP) mode (plasmon for a metal, phonon for an ionic crystal in the Restsrahlen band, Zenneck wave for other lossy materials \cite{wait1998}...) and a "photonic" field  ressembling the dipole-type field (\ref{Hperf}) of the perfect metal case, say:
\begin{equation} H_z(x,y>0)=H^{SP}_z(x,y)+H^{Ph}_z(x,y). \end{equation}
The proper pole of the integrand corresponds to the tranverse plasma oscillation generated by each metallic edge of the aperture. Applying the residue theorem, for $y>0$,
\begin{equation} 
\label{Hsp} H^{SP}_z(x,y)= \alpha \frac{2i\pi Z}{n_{SP}} S_{w,Z}(n_{SP})\  e^{i k( n_{SP}|x|-Zy)},
\end{equation}
with $n_{SP}=\sqrt{1-Z^2}$ ($\Re(n_{SP})>1$). For $|x|<w/2$, $H^{SP}$ still has a plasmonic nature since it is supported by the guided mode of the cavity. Eq.$(\ref{Hsp})$ (not really new) explains the trade-off on the ratio $\Im(\varepsilon)/\Re(\varepsilon)$ to generate a strong and long-range surface polariton mode. The aperture function $S_w(n_{SP})$ also implies that when $w\approx\lambda/\Re(n_{SP})$, destructive interferences annihilate the polariton whatever the slit reaction (see Fig\ref{figure1}(b)). However, for some materials with $\Im(\varepsilon) \sim -\Re(\varepsilon)$, the cutoff-width condition is never fully fulfilled (see the amorphous silicon case) and a weak surface polariton can always be launched from the slit. On the other side, when $w \ll \lambda $, the waveguide index $n_g$ exhibits an increasing imaginary part (absorption) which may attenuate the $\alpha$ coefficient, and then, the SP generation, as experimentally observed\cite{lalanne2009}.

\begin{figure}[h]
\includegraphics[angle=0,scale=0.29]{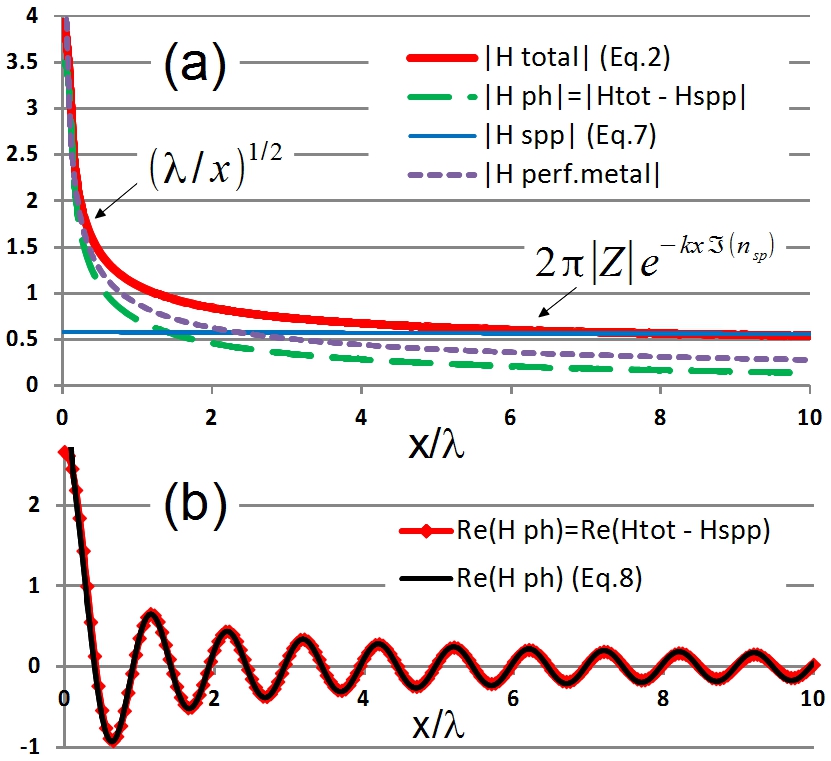}
\caption{Magnetic field scattered on a gold surface by a slit $400nm$ width at $\lambda=1.5\mu m$. (a) field modulus, (b) real part. $H_{total}$ is the scattering integral (\ref{Htot}) taking $\alpha=1$.\label{figure2}}
\end{figure}

Before going ahead, Figure \ref{figure2} gives a numerical example showing the weight of each magnetic component scattered over a gold surface at $\lambda=1.5\mu$m ($w=400$nm). While the plasmon and the photonic field are both proportional to the coefficient $\alpha$, $H^{Ph}$ predominates in the neighbouring of the sub-$\lambda$ aperture, with a damping similar to that of the perfect metal case. Due to different spatial damping, the polariton is rapidly the majority mode over many wavelengths. This is a general behaviour provided that $|Z|$ and/or $|S_w|$ (Fig.\ref{figure1}(b)) are not too small. 

However, the underlying physics is far from being complete. Indeed, on the basis of Ref.\cite{dai2009}, one can show that the photonic field near the slit may be expressed analytically thanks to a fine correction of the field of the perfect metal case (numerically verified in Fig.\ref{figure2}(b)). While $x|Z|/\lambda <1$, and for $x>w/2$, the real case gives:

\begin{eqnarray}\label{HphZ}
H_z^{Ph}(x,y=0) \approx  \alpha \sec (\frac{kw}{2})\cdot  (\sqrt{\frac{2\pi}{kx}}e^{-i\frac{\pi}{4}}-i\pi Z \nonumber \\
- Z^2\sqrt{2i\pi kx}-\frac{\pi kx}{2} Z^3) \cdot \ e^{ikx} \label{HphZ}
\end{eqnarray}

The correction $\sim i\pi Z$ seems related to some energy transfer to the surface polariton, when we compare it to Eq.(\ref{Hsp}). The  propagating terms $\sim Z^2,Z^3$ are stranger as they are not linked to absorption losses, but to small out-of-phase \textit{radiations}.
Actually, as we will detail hereafter, in the real metal case, light waves come from at least \textit{two} contributions, that is to say
$H^{Ph}_z(kx>0)=\alpha  \sqrt{2\pi/kx}\ F(kx)e^{ikx} $, where $F(kx)$ is an enveloppe function comprising always the dipolar (cylindrical) field, but hiding another radiating subfield which is \textit{not} cylindrical. To show this, not only at the surface level, but in the whole space, a relevant approach is to study the asymptotical behaviour of the field far enough from the aperture. This is the subject of the following section.

\section{Scattered photonic field: analytical Norton and dipole contributions}

To find the asymptotical behaviour of the far field, a way consists in resorting to a double second-order Taylor expansion. 
Indeed, let us consider again the integral (\ref{Htot}). One may put the phase $\phi(u)=k(ux+vy)$ and $f(u)=S_{w,Z}(u)/(v+Z)$.
We will apply the stationnary phase method for $(Z,y)\neq (0,0)$.  It can be intuitively understood that, although the field results from the contribution of a whole continuum of wave vectors, the oscillations of the exponential become extremely rapid at large distance, with destructive interference of the spectrum, except when the phase $\phi(u)$ is nearly constant, close to an extremum. The condition $\phi^{'}(u)=0$ is indeed fulfilled for a unique wave vector $(u,v)=(u_0,v_0)=(x/r,y/r)$, which corresponds to a radiated field. Thus, around $u_0=x/r$:
\begin{equation}
\phi \approx\phi(u_0)+\frac{(u-u_0)^2}{2}\phi^{''}(u_0)= kr[1-\frac{(\Delta u)^2}{2}(\frac{r}{y})^2].
\end{equation}

A Taylor expansion of $f(u)$ is also applied, noticing that the first order term $\propto f^{'}(u_0)$ will be null.
We are then driven to calculate different Fresnel integrals. If we introduce polar coordinates, by naturally putting $(u_0,v_0)=(\sin\varphi,\cos\varphi) $, with $\varphi \in [-\pi/2;\pi/2]$ an angle defined with respect to the $y$ axis, we finally get:
\begin{equation}\label{Hphot}  H_z ^{Ph}(kr \gg 1) = H_z^{Dip}(kr)+ H_z^{N}(kr)+O(kr^{-5/2}),\end{equation}
with 
\begin{equation}\label{Hdip}  H_z^{Dip}(kr,\varphi) =  \alpha S_{w,Z}(\sin\varphi) \frac{\cos\varphi }{\cos\varphi + Z} \sqrt{\frac{2\pi}{kr}} e^{i(kr-\frac{\pi}{4})} , \end{equation}
\begin{equation}\label{Hn} H_z^{N}(kr,\varphi) = \alpha \ \Theta(\varphi,Z)\ \sqrt{\frac{2\pi}{(kr)^3}}  e^{i(kr+\frac{\pi}{4})} , \end{equation}
and
\begin{equation}\label{Theta} \Theta(\varphi,Z ) \approx \frac{S_{w,Z}(\sin\varphi)}{ (Z+\cos \varphi )^2} [1+ \frac{2\cos\varphi\sin^2\varphi }{Z+\cos \varphi }] . \end{equation}
High order terms or minor ones (see Appendix B) are neglected. 
Through several simulations, we find a typical validity threshold $x/\lambda>10/|Z|$ over which the asymptotical expression of $H_z^{N}$ starts to fit with the numerically calculated field, at the surface.

Let us get insight into both electromagnetic entities obtained. The first contribution $H_z^{Dip}$ is the conventional dipolar field, which is actually valid whatever $(Z,\varphi )$. We analytically see that when $Z=0$, we retrieve the perfect metal case with a far-field persisting at the surface ($\varphi =\pi/2$) with a $1/\sqrt{kr}$ damping. But when $Z\neq 0$, $H_z^{Dip}(\varphi \sim \pi/2)=0$: we have the appearance of a shadow zone \cite{nikitin2009} (adjacent to the surface) for the radiated power, leaving  only the place to the surface polariton and $H_z^{N}$, when we are sufficiently far from the cavity. 

\begin{figure}[h]
\includegraphics[angle=0,scale=0.17]{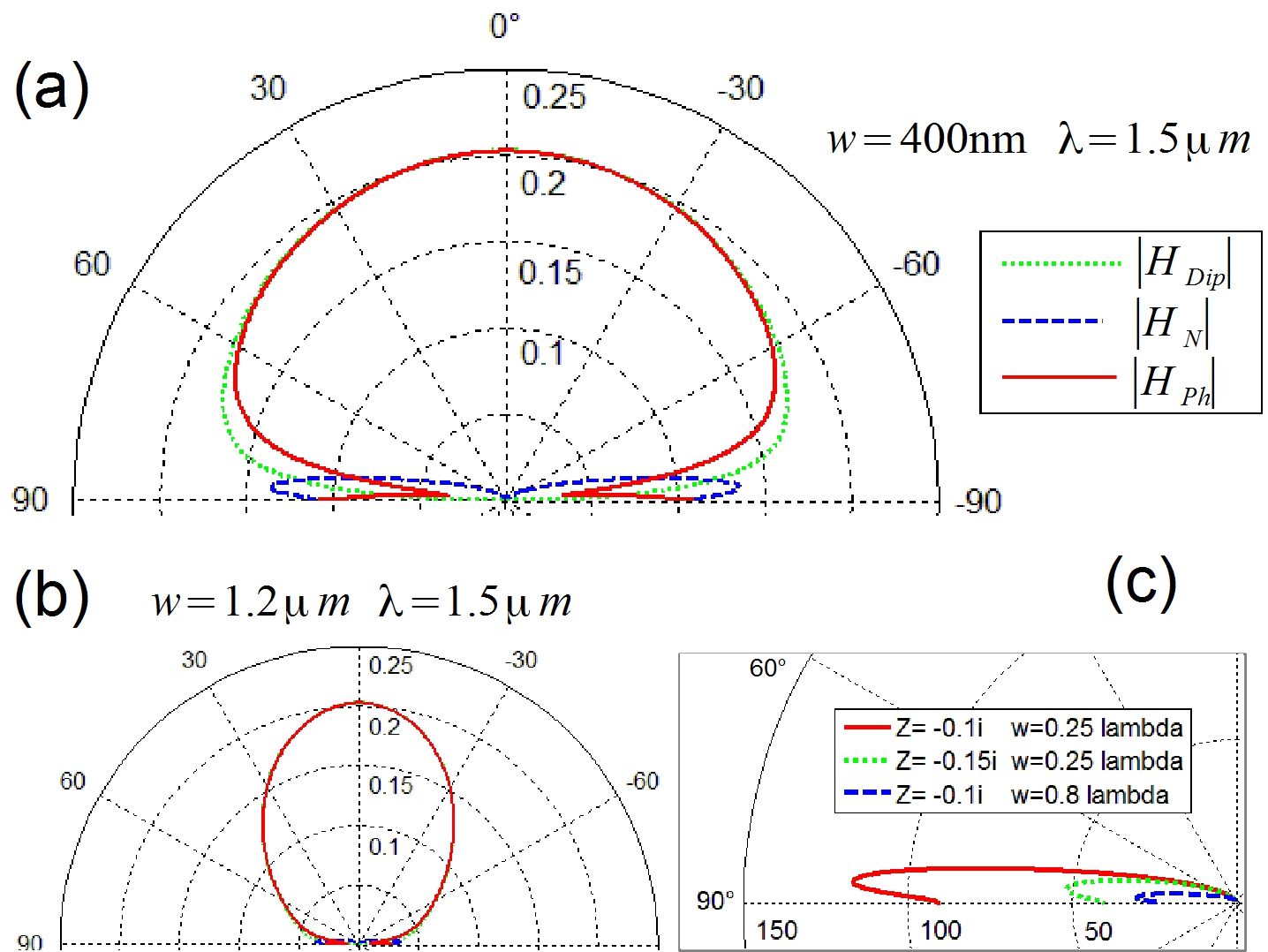}
\caption{Polar behaviour of the analytical expressions $H^{Dip}$ (\ref{Hdip}), $H^N$ (\ref{Hn}), and of the sum $ H^{Ph}$ (\ref{Hphot}) taking gold at $\lambda=1.5\mu m$ ($Z=0.0058-0.1048i$) and $kr=150$: (a) $w=0.4\mu$m (b) $w=1.2 \mu$m. (c) $|\Theta(\varphi)|$ (\ref{Theta}) for some values of $Z$ or $w/\lambda$.
\label{figure3}}\end{figure}  

The second one, noted $H_z^{N}$, is what has a direct link with the so-called Norton wave. This light field is not a real surface one, although spatially concentrated due to its rapid decrease in radial amplitude. The radiation pattern $\Theta(\varphi)$ is a meaningful result as it will indicate the physical origin of these additional lightwaves.  For grazing angles, it is not null at the surface but proportional to $1/Z^2=\varepsilon$, which could be high if it was not counterbalanced by the $r^{-3/2}$ damping. Other calculations\cite{lalanne2009,nikitin2009,dai2009}, made for a punctual scatterer, seem to be consistent with this $\varepsilon$ factor, but this has not been commented. Besides, $H_z^{N}$ presents an intrinsic phase quadrature with respect to $H_z^{Dip}$, hence possible destructive interferences of both fields in polar directions where they present close amplitudes. Although $H^N$ is not unambiguously given for small $kr$, Eq.(\ref{HphZ}) would be an indication that $H^N$ is likely present and immersed into the preponderant dipole field. It is worth noting that, by strictly keeping cartesian coordinates and taking $ r\approx x$ close to the surface, one gets a $y/(x^{3/2})$ behaviour \cite{gravel2012} for $H_z^{Dip}(kr)$ but this is misleading: the true wave which is non-null at the surface with a $1/r^{3/2}$ damping  is $H_z^{N}(kr)$. We recall that the fields described here are generated by a scatterer that extends infinitely in one of the spatial dimensions. For a \textit{point} scatterer, like in the well-known Sommerfeld problem\cite{wait1998}, the field of a surface plasmon (or a Zenneck surface mode) behaves as $1/\sqrt{r}$ whereas the Norton wave is as $1/r^2$. 

\begin{figure}[h]
\includegraphics[angle=0,scale=0.26]{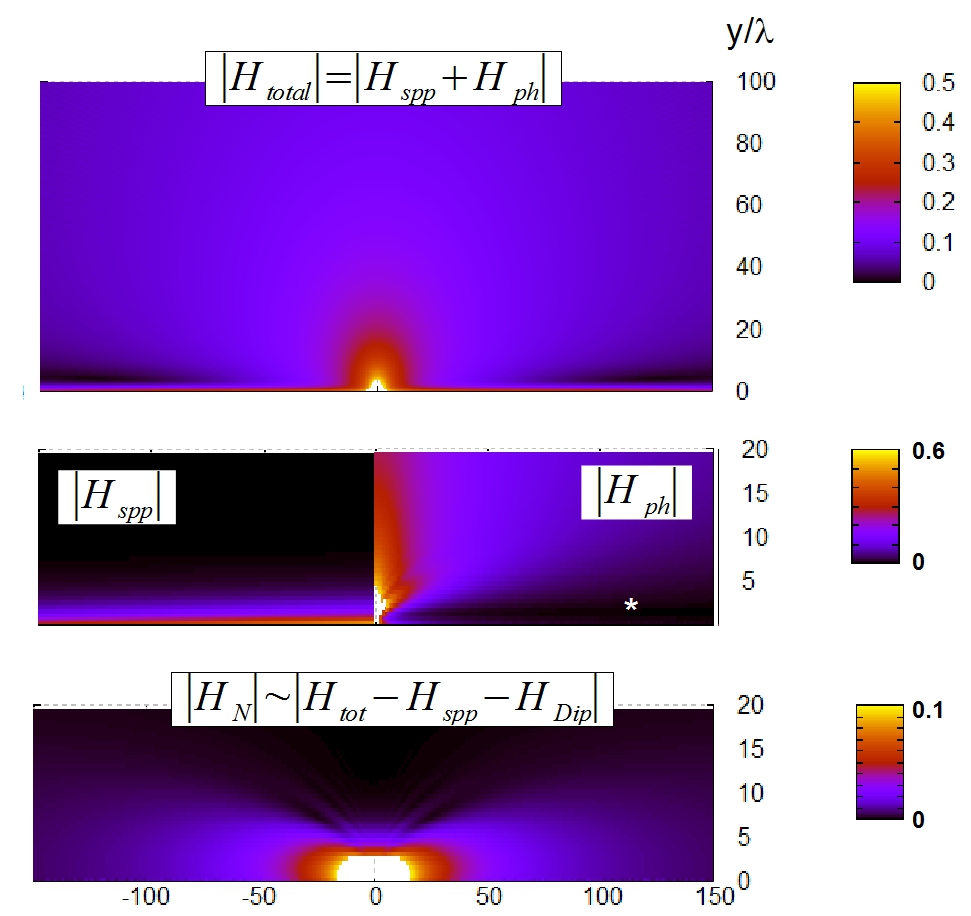}
\caption{Maps of field modulus for a slit width $w=400nm$ and gold at $\lambda=1.5\mu m$: $H^{total}$ results from the exact integral (\ref{Htot}) taking $\alpha=1$, $H^{spp}$ is (\ref{Hsp}), $H_z^{Ph}$ is (\ref{Htot})-(\ref{Hsp}), and $H_z^{N}$ results here from (\ref{Htot})- (\ref{Hsp})-(\ref{Hdip}). Let us recall that calculus of $H_z^{N}$ through this subtraction is not valid close to the aperture. As $H^{total}=4$ just above the aperture, max of scales are willingly limited to better highlight the details of the scattering patterns. The white star refers to a spatial zone where various field components are detailed in Fig.\ref{figure5}.\label{figure4}}
\end{figure}

Figure \ref{figure3}(a) shows the behaviour of the total light field $H_z^{Ph}$ and that of its inner components. Unsurprisingly, $H_z^{Dip}$ is quasi-isotropic, typical of a Rayleigh scattering. $H_z^{N}$ is clearly different and presents two horizontal half-lobes at each side of the cavity, which are physically \textit{connected} to the metallic surface. But the remarkable effect is that combination of $H_z^{Dip}$ and $H_z^{N}$ strongly modifies the final radiation pattern of $H_z^{Ph}$: it \textit{gives the impression} that a residual surface wave slides along the metal ($y=0$), that is not of plasmonic nature. We also understand here that the \textit{transition} from a $1/ \sqrt{x}$ damping (near the cavity) to a $1/ \sqrt{x^3}$ (far away from it) at the surface, already observed \cite{lalanne2009}, is due to \textit{a change of the major scattering contribution} (from the dipole/cylindrical field towards the Norton field), without change of the surface wavevector $k$. Purely numerical simulations cannot allow us to explain such a continuous transition, without analytical developments. Outside the shadow zone, i.e. quite above the surface, we find the preponderant dipolar space field. 

Other remarks can be made: as exemplified in Fig.\ref{figure3}(b), when $w\rightarrow\lambda$, $H_z^{N}$ vanishes (the surface polariton vanishes even more, according to Fig.\ref{figure1}(b)) and the radiation lobe of $H^{Ph}$ becomes more focused in the normal direction, which may present some interest to transmit light in a less dispersed beam. The behaviour of $\Theta(\varphi )$ for some arbitrary values of $Z$ or $w/\lambda$ is also given in Fig.\ref{figure3}(c). When $Z\rightarrow 0$, the semi-lobes of $H^{N}$ remain flattened against the surface, and the polar angle corresponding to their maximum amplitude tends to $\pi /2$. The space of validity of (\ref{Hn}) is also rejected to infinity: at end, only $H_z^{Dip}$ becomes relevant, and we retrieve the perfect metal case. 

\begin{figure}[h]
\includegraphics[angle=0,scale=0.29]{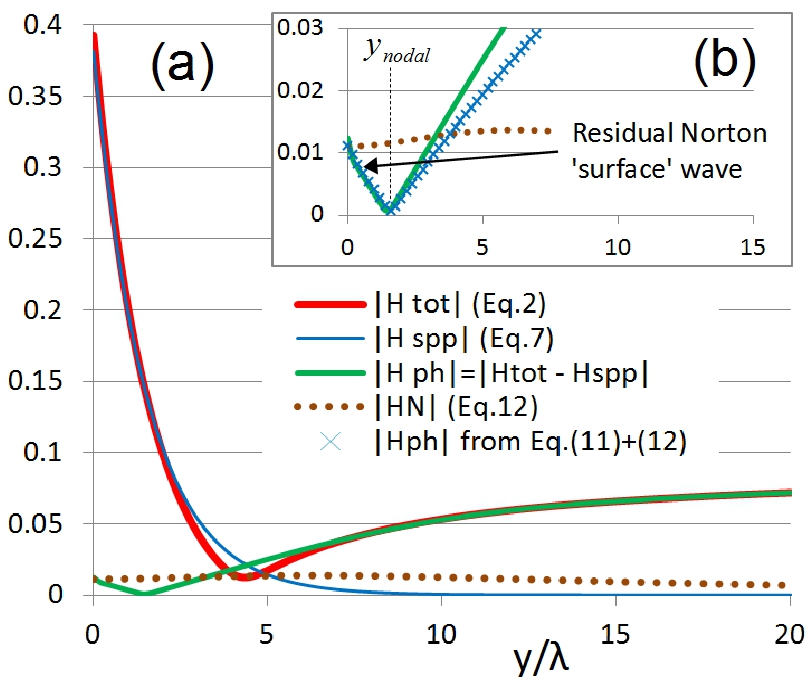}
\caption{(a) Behaviour of the magnetic field perpendicularly to the gold surface, far enough from the cavity at $x/\lambda=115$. This spatial region is indicated by a white star in the $H_z^{Ph}$ map of Figure \ref{figure4}. The plasmonic and photonic contributions are also superposed. The photonic field comprises the dipolar field (null at the surface) and the Norton one. Insert (b) gives details of the photonic part near the surface (numerically and analytically calculated). One can show by hand that $y_{nodal}\approx 1/(-k\Im (Z))$ is the ordinate of a plane along which $H_z^{Ph}$ cancels, whatever $x$ (for large $kr$). \label{figure5}}
\end{figure} 

\section{Physical interpretation of the Norton field}

What is the physical source of $H_z^{N}$? If we remind the $E_y$ arrows in Fig.\ref{figure1}(a), we have an effective \textit{vertical} electric dipole at the aperture level (see also Appendix A). This vertical momentum is responsible for the Norton wave generation, whereas the \textit{horizontal} momentum $E_x$, at the mouth of the aperture, forms the classic oscillating dipole (hence the phase quadrature between (\ref{Hdip}) and (\ref{Hn})). The electromagnetic radiation $H_z^{N}$ is linked to induced surface currents occuring in the skin-depth of both metallic parts forming the slit.  Indeed, the cavity can be viewed as a capacity under illumination whereas horizontal metallic parts play the role of (dissipative) inductances \cite{Yang2011}. The current component
\begin{eqnarray}
J_x (kx\gg 1)= \int_{-\infty}^{0} H_z^N(x,y) dy=\frac{Z}{ik} H^N_z (x,y=0) \ \ 
\label{current}
\end{eqnarray}
which is guided along the surface corresponds to an oscillating field having the wave vector of the free space ($k$). Thus, while $\alpha\neq 0$, the \textit{surface itself radiates} (infrared waves, in our example) and behaves as a uniform \textit{leaky-wave-antenna} \cite{Jackson2012}. The surface polariton has nothing to do with it, but simply superposes to the Norton field, and does not radiate. Thus, a great difference is that $H_z^{sp}$ is damped due to absorption (see Eq.\ref{Hsp})) whereas $H_z^{N}$ is dissipated through emission. They have not the same $Z$-dependence. The $\varepsilon$ amplitude of $\Theta$, when $\varphi \approx\pi/2$, seems related to a limit of the spreading of the surfacic charge around the slit (according to the metal conductivity). Also, as observed in Fig.\ref{figure3}, there is a grazing angle $\varphi_m$ for which $|\Theta (\varphi)|$ meets a maximum. This angle is close to the pseudo-Brewster angle defined for metals, that is to say the emissivity of the radiating surface is maximum for polar directions where its reflectivity should be minimum (in TM-polarization).

Figure \ref{figure4} gathers some  fully calculated maps of the total scattered field (\ref{Htot}) and its inner components. The first map clearly shows the conventional dipole-type radiation, the surface polariton, and the presence of a shadow zone. While $H^{Ph}$ reveals a non-negligible amplitude near the aperture, the existence of a radiating field $H^{N}$ localized along the whole surface is confirmed by simply subtracting the polariton (\ref{Hsp}) and the dipolar field (\ref{Hdip}) from the total field. The new map exhibits the scattering lobes with a butterfly shape predicted by the $\Theta$ function. \textit{This physical picture has gone unnoticed in the literature devoted to optics of metallic nano-structures}. Additional details of the photonic component near the surface, far from the slit, are displayed in Fig.\ref{figure5}. Along a direction normal to the surface, it well presents a small amplitude at the metal level (same value given by (\ref{Hn}) taking into account the finite size $w$), cancels rapidly at the nodal plane $y= [-k\Im (Z)]^{-1}$, and increases again when one progressively enters the dipolar lobe, as expected from Fig.\ref{figure3}(a). Whereas the true Norton field extends over a few wavelengths in the $y$ direction, its combination with the dipole field reduces to a virtual 'surface' wave. However, the plasmon wave remains overwhelming at the interface, due to its low damping on gold surface in the mid-infrared.

\section{Case of metal-like, strongly absorbing materials}
	
\begin{figure}[h]
\includegraphics[angle=0,scale=0.33]{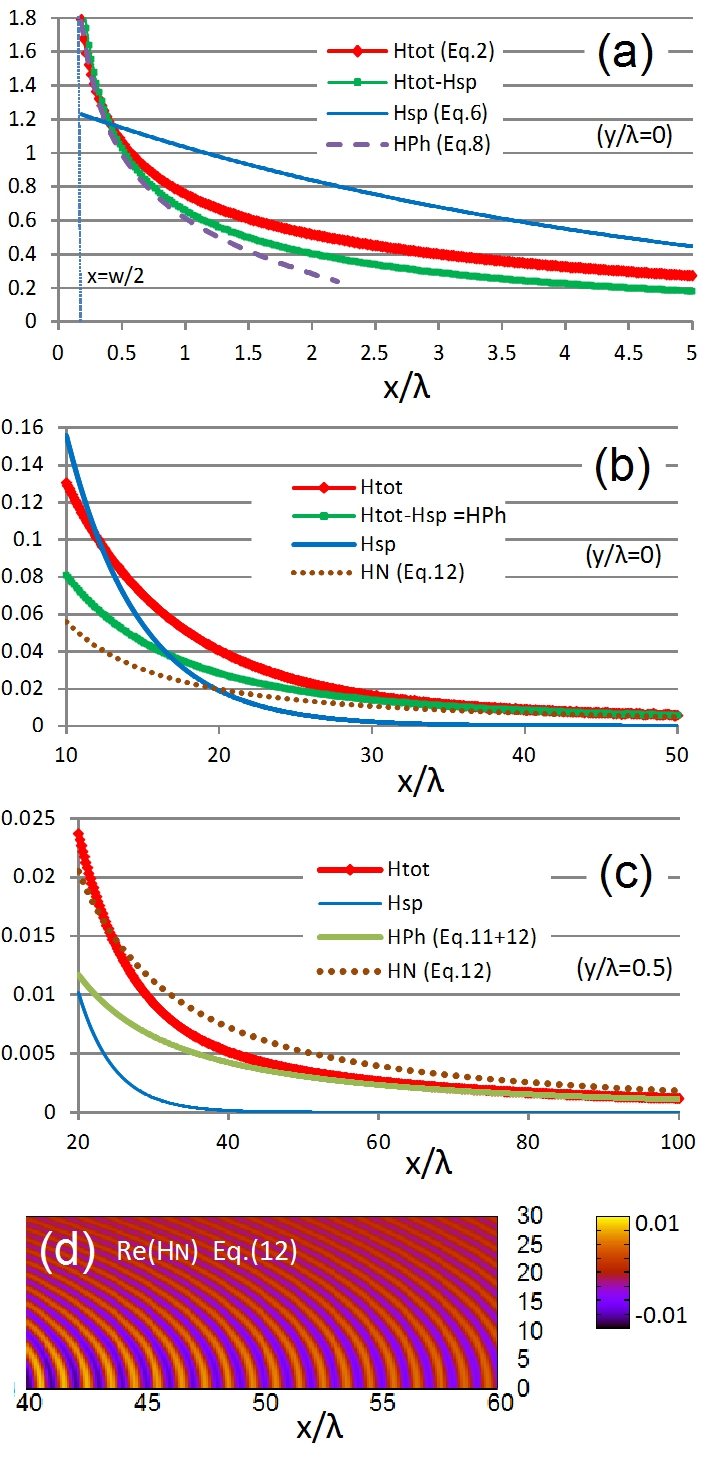}
\caption{Example given with amorphous silicon at $\lambda=282$nm, with $\varepsilon_{aSi}=-3.7+13.8i$ \cite{palik} ($|Z|=0.26$) and $w=100nm$. Modulus of the total field and that of its plasmonic and radiative components: (a) near the slit (for $y=0$), (b) far from the slit at the surface level, and (c) far from the slit above the silicon at $y=\lambda/2$. We check that the analytical expression (\ref{Hn}) of $H^{N}$ at the surface well reflects the numerically calculated lightfield while $x/\lambda>10/|Z|$. (d) Real part of $H_z^{N}$.\label{figure6}}
\end{figure} 

On an experimental level, the Norton-type wave will be really apparent provided that the SP collapses, i.e. faraway from the source, or for strongly absorbing materials. 
So, instead of gold, let us consider the case of amorphous silicon in the ultraviolet region. Indeed, $aSi$ behaves as a metal and shows strong absorbing properties in this spectral domain. 
Figures \ref{figure6}(a) and (b) display the amplitudes of the field scattered at the surface, for two ranges of distance, $w/2<x/\lambda<5$ and $10<x/\lambda<50$ ($w=100$nm). The total field (\ref{Htot}) is compared to analytical expressions of each physical contribution (in their domain of validity). Near the slit, the light field is always predominant. When $x/\lambda>2/5$, the SPP becomes the stronger contribution but the total field keeps a lower amplitude. Above $x/\lambda=12$, the total field presents an amplitude stronger than $H^{SPP}$. Finally, very far from the slit (see Fig. \ref{figure6}(b)), for $x/\lambda>35$, the polariton completely vanishes whereas the Norton-field remains dominant (well described by Eq.(\ref{Hn})). This is all the more notable when we evaluate $H_z$ a little bit above the surface, at $y/\lambda=0.5$ for instance, as shown in Fig.\ref{figure6}(c). Clearly, $H_z^N$ may present an amplitude greater than that of the total field. At $y/\lambda=5$ (not shown), the SP field is negligible, the dipolar field is prevalent, and the Norton field still exists but with a weaker amplitude. 

The chosen example, here, may suggest a challenging experiment to better observe and quantify the predicted wave (instead of a gold surface), with a wavefront sensor\cite{Bon2012} for instance.
Let us note that, if the surface impedance model is less rigorous in the visible, with moderate permittivity moduli, we can meet the same behaviour, in the infrared region, with other materials like Titanium Nitride (TiN) which behaves as an absorbing metal (for example, $\varepsilon_{TiN}=-18.2+27.1i$ at $\lambda=1.24\mu$m \cite{palik}, and a special demonstration at telecom wavelength could be welcome). The scattered surface fields should be preferably probed at the opposite side of the slit, in order to omit the interference of the exciting (incident) wave. What is more, to enhance the field amplitudes  (i.e. $\alpha$), the slit height should be roughly $\lambda/2$ to generate a well-known Fabry-Perot resonance. Instead of a narrow slit, a patch antenna (horizontal cavity) could also launch Norton and plasmon waves, as the lateral apertures are the sites of vertical electrical fields. 

\section{Conclusion}

As a conclusion, this paper brings comprehensive analytical and numerical results which shed light on a particular kind of
electromagnetic wave scattered by a narrow slit on metal (or highly conducting) surfaces. Its physical source and
behaviour, in the broad space and at optical frequencies, had not been clearly identified in the
scientific literature until now. Indeed, the well-known surface plasmon and the classic dipolar light field are not the only contributions. A lateral Norton-type wave also exists, most often immersed within the scattered field. This is \textit{not} a cylindrical field, neither a true surface wave, nor a proper mode of a flat metallic interface, but a reaction to a polarized excitation: it features a special radiation pattern ($\Theta$ function) taking its origin into induced surface currents residing in the metal skin depth and generated by the dipole aperture (excited by the vertical electrical momentum). 

Considering noble metals, this field might generally have too weak absolute amplitude to be practically exploited in systems of photonic size, but depending on other frequencies and permittivities (as illustrated above, in the paper), it may become a relevant wave to convey information faraway at interfaces (not necessarily flat). For example, such Norton waves are expected to be the best candidate to transmit microwave signals on the (conducting) human skin, because Zenneck-surface modes are loosely excited\cite{Lea2009}.

Although the present work does not aim at revisiting the extraordinary optical transmission (EOT) (taken as a starting context), an open question could be to know to which extent \cite{note} radiation from induced currents may be channeled through diffraction orders and participate to EOT. Indeed, in some cases, EOT cannot be assisted by plamons, but by other vectors with materials whose permittivity exhibits an imaginary part and a positive real part\cite{Sarrasin2005}. One can also wonder if there is an equivalent of the Norton-type surface radiance in acoustics \cite{Lu2007}. 
Thus, this deeper fundamental understanding of the canonical slit case, that bridges the gap between nano-optics and leaky-wave antennas, may inspire further investigations and other ways of lightwave engineering, with structures more elaborate.

\appendix

\section{Scattered surface fields in the perfect metal case}

This section gives some analytical expressions of the electromagnetic fields scattered at the surface ($y=0$) by a narrow one-dimensional slit (supporting a fundamental guided mode), in the case of the perfect reflector. Results are based on the exact integral Eq.(\ref{Htot}).

\subsection{Perpendicular electric field $E_y$}

As $E_y = (1/ik) \partial H_z/ \partial x$ (excepting a $c\varepsilon_0$ factor), this component may be expressed for all $x$, and $\forall y>0$ as:

\begin{eqnarray}
 E_y^{scat}(x,y)=\frac{\alpha\pi}{ikw}[H^{(1)}_0(k\sqrt{(x+w/2)^2+y^2}) \nonumber \\
 -H^{(1)}_0(k\sqrt{(x-w/2)^2+y^2})] 
\end{eqnarray}
\[\mbox{where}\quad H^{(1)}_0(k\sqrt{x^2+y^2})=\int_{-\infty}^{+\infty} \frac{e^{ik(ux+v|y|)}}{\pi v}du , \]
and $v=\sqrt{1-u^2}$. $H^{(1)}_0$ is the zero order Hankel function of the first kind. Given the angular spectrum representation, it is interesting to separate the respective contributions due to \textit{evanescent} waves ($|u|>1$) from that due to \textit{space} waves ($|u|<1$), so that we can write $E_y=E_y^{eva}+E_y^{spa}$.
Thus, at the interface,  and $\forall x$:
\begin{equation}
E_y^{eva}(x,y=0)= \frac{\alpha\pi}{kw}[Y_0(k|x+w/2|)-Y_0(k|x-w/2|)] 
\end{equation}
where $Y_0(a)$ is the Bessel function of the second kind (also noted $N_0(a)$ in the literature), defined in the sense that $a>0$.
For the homogeneous contribution, we have:
\begin{equation} E_y^{spa} (x,y=0)= \frac{\alpha\pi}{ikw}[J_0(k(x+w/2))-J_0(k(x-w/2))]
\end{equation}
where $J_0(a)$ is the Bessel function of the first kind. 

To clarify our understanding, the above analytical expressions have the following asymptotical behaviours:
\begin{eqnarray} E_y^{spa}(x,y=0)= \left \{ \begin{array}{lll}i\alpha\frac{\pi}{2} kx \quad \mbox{if $|x|<\frac{w}{2}$}\\
-i\alpha\sqrt{\frac{2\pi}{kx}} \cos(kx+\frac{\pi}{4}) \mbox{ else}
\end{array} \right. 
\end{eqnarray}
\begin{eqnarray} E_y^{eva}(x,y=0)=\left \{ \begin{array}{lll}
(\frac{2\alpha}{kw})\ln \left[\frac{x+w/2}{w/2-x}\right] \quad \mbox{if $|x|<\frac{w}{2}$}\\
\alpha\sqrt{\frac{2\pi}{kx}}\sin(kx+\frac{\pi}{4})  \mbox{ if $x>\frac{w}{2}$}
\end{array} \right. \end{eqnarray}

\begin{figure}[h]
\includegraphics[angle=0,scale=.27]{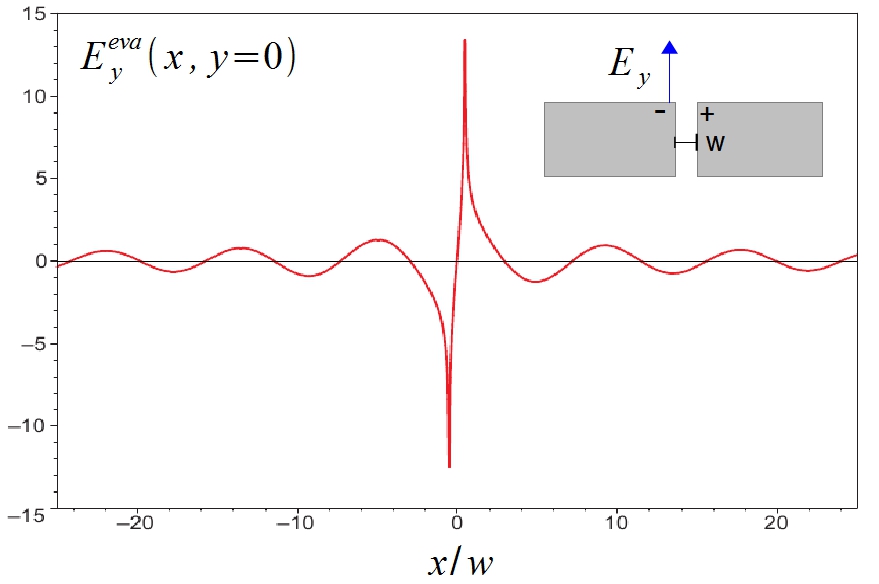}
\caption{Exact behaviour of $E_y^{eva}$ along the interface $y=0$, in the case of the perfect metal, depending on the reduced variable $x/w$, for $kw=0.75$. The slit, centered at $x=0$, occupies the interval $[-\frac{1}{2};\frac{1}{2}]$. The excitation coefficient $\alpha$ of the cavity mode has been fixed to unity here.}\label{fig:ey}
\end{figure}

We note that, in the case of a perfect metal, the \textit{evanescent} contribution $E_y^{eva}$ has an intrinsic singularity at the corners $x=\pm w/2$, which is inherent in a point effect in this polarization. This solution is physically justified since, above the aperture, the field behaves as $\ln(w/2-x)$ when $x\rightarrow w/2^{-}$, so it is no more singular than $1/\sqrt{w/2-x}$: this joins the Meixner edge condition that ensures the field is bounded in energy \cite{bolomay71}. Let us emphasize that $E_y\propto \sigma_{hor}$ where $\sigma_{hor}$ is the surface charge density on the horizontal interfaces. As an example, Fig.~\ref{fig:ey} displays the behaviour of $E_y$ (exact analytical expression) taking $\alpha=1$, and shows that this component is solely intense around the slit edges: such sites constitue the hot spots where the electric charges strongly accumulate if a resonance occurs.

\subsection{Magnetic field $H_z$ at the surface}
For the scattered magnetic field, we have, for $y=0$:

\begin{equation} H_z(x,y)=\frac{2\alpha}{kw}\int_{0}^{\infty} \frac{\sin(k(\frac{w}{2}+x)u)-\sin(k(x-\frac{w}{2})u)}{u\sqrt{1-u^2}}du
\end{equation}
We may judiciously rewrite the integral as follows:
\begin{equation} H_z(x,y=0)=\frac{\pi\alpha}{kw} \int_{k(x-\frac{w}{2})}^{k(x+\frac{w}{2})} [ J_0(t)+iY_0(t) ] dt.
\end{equation}
The magnetic field results from the sum of a continuum of line sources all over the aperture width (according to the Huygens principle). By separating the contribution due to evanescent waves ($|u|>1$) from that due to space ones ($|u|<1$), in the first integral, 
we can write $H_z(x,y=0)=H_z^{spa}+H_z^{eva}$. Both contributions have the respective asymptotical behaviours:
\begin{eqnarray}
H_z^{spa}|_{x,y=0}= \left \{ \begin{array}{ll}
\pi \alpha (1-(kx/2)^2)\quad \mbox{if $|x|<\frac{w}{2}$}\\
 \alpha\sqrt{\frac{2\pi}{kx}} \sin(kx+\frac{\pi}{4}) \quad \mbox{for $x>\frac{w}{2}$}
\end{array} \right. \end{eqnarray}
\begin{eqnarray}
H_z^{eva}|_{x,y=0}=\left \{ \begin{array}{ll}
\frac{2\alpha}{i}[1+2\ln(2)-\ln(kw)-\gamma] \mbox{ if $|x|<\frac{w}{2}$}\\
\frac{\alpha}{i}\sqrt{\frac{2\pi}{kx}} \cos(kx+\frac{\pi}{4})\quad \mbox{for $x>\frac{w}{2}$}
\end{array} \right. \end{eqnarray}
where $\gamma=0.55721...$ is the \textit{Euler}'s constant. 

\subsection{Tangential electric field $E_x$ at the surface}

$E_x = (i/k) \partial H_z/ \partial y$. Here again, we separate the evanescent and space wave contributions:

\begin{eqnarray}
E_x^{eva}|_{x,y=0}=\left | \begin{array}{lllll}
\frac{2\alpha}{kw}[Si(k(x+w/2))-Si(k(x-w/2))]  \\
\quad\quad\quad \mbox{ if $|x|>w/2$} \\
\\
-\frac{2\alpha}{kw}[\pi-Si(k(x+\frac{w}{2}))+Si(k(x-\frac{w}{2}))] \\
 \quad\quad\quad \mbox{ if $|x|<w/2$}\\
\end{array} \right.
 \end{eqnarray}
and, $\forall x$:
 \begin{equation}
 E_x^{spa}|_{x,y=0}= -\frac{2\alpha}{kw}[Si(k(x+\frac{w}{2}))-Si(k(x-\frac{w}{2}))] 
\end{equation}
where $Si(a)$ designates the Sine Integral defined by:
\[Si(a)=\int_0^a \frac{\sin t}{t}dt\]

By summing $E_x^{spa}+E_x^{eva}$, one actually finds that $E_x=0$ on the perfectly reflecting surface (as expected), except along the sub-wavelength aperture:
\begin{equation} E_x(|x|<\frac{w}{2},y=0)=-\frac{\alpha\lambda}{w} \end{equation}
This is the amplitude of the horizontal dipolar momentum existing at the slit output (such a dipole oscillates in time, hence radiation). Indeed, $E_x(x=\pm w/2,y)\propto \sigma_{vert}$ where $\sigma_{vert}$ is the surface charge density on the vertical walls, the charge distributions on both sides of the slit having opposite signs.

 It is possible to show  theoretically\cite{leperchec2011} that  the normalized field modulus $|E_x/E_{inc}|$ cannot overcome $2/kw$ at the mouth of a transmitting sub-wavelength slit, in the perfect metal case, i.e. $|\alpha| \leq 1/\pi$.\\

\section{Corrective terms of $\Theta$}

When we calculate the second derivative of $f(u)=S_{w,Z}(u)/(v+Z)$, we find a main term for the $\Theta$ function (already given in Eq.(\ref{Theta})), and other minority terms linked to the aperture size dependence (regardless of whether the metal is perfect or not). These last ones can be neglected to simplify our analysis and were not essential for a quantitative description. Both complementary contributions of $\Theta$ will be called $\Theta^{compl,1}$ and $\Theta^{compl,2}$.
\begin{equation} 
\Theta^{compl,1}(\varphi,Z ) = 2 \frac{d S_{w,Z}(u)}{d u}|_{u=\sin\varphi} \frac{\sin\varphi \cos^2 \varphi}{(Z+\cos \varphi )^2}. 
\end{equation} 

Assuming, for sake of simplicity, that $S_{w,Z}(u)=\sec(kwu/2)$ as in the perfect metal case, we get: 

\begin{equation} \label{Thetacomp} 
\Theta^{compl,1}(\varphi,Z ) \approx 2 [\cos(\kappa)-\sec(\kappa)] [\frac{\cos \varphi}{Z+\cos \varphi}]^2. 
\end{equation}

with $\kappa=(kw/2)\sin\varphi$. As $\cos(\kappa)-\sec(\kappa)\approx -\kappa^2 /3$ for small $\kappa$, $\Theta^{compl,1}$ has only some numerical weight for angles $\varphi \sim \pi/3$, and it is zero when $\varphi=0$ or at the surface level. It would describe some \textit{quadrupolar}-order scattering lobe of the slit. It is a geometrical aperture-term as $\Theta^{compl,1}$ does not cancel when $Z=0$ and tends to vanish when $w\rightarrow 0$. Also:

\begin{equation} 
\Theta^{compl,2}(\varphi,Z ) =  \frac{d^2 S_{w,Z}(u)}{d^2 u}|_{u=\sin\varphi} \frac{\cos^3 \varphi}{Z+\cos \varphi}. 
\end{equation} 

\begin{equation}
 \frac{d^2 S_{w,Z}(u)}{d^2 u}|_{u=\sin\varphi}=\frac{2[\sec(\kappa)-\cos(\kappa)]-\kappa^2\sec(\kappa)}{\sin^2\varphi}
\end{equation} 

Note that $\Theta^{compl,2}/\Theta^{compl,1} \sim Z+\cos \varphi$, with $|Z|< 1$.

These small lacking corrections have no contribution at the surface, that is why they are not directly included in the Norton field expression $H^{N}$.

\end{document}